 \documentclass[5p,times]{elsarticle}

\usepackage{amssymb}
 \usepackage{amsmath} 
\usepackage{xcolor}
\usepackage{soul}
\usepackage{listings}
\definecolor{comment-text-color}{rgb}{0,0.8,0.6}

\journal{Future Generation Computer Systems}

\begin{document}
\tnotetext[t1]{This manuscript has been authored by UT-Battelle, LLC, under Contract No.~DE-AC0500OR22725 with the U.S.~Department of Energy. The United States Government retains and the publisher, by accepting the article for publication, acknowledges that the United States Government retains a non-exclusive, paid-up, irrevocable, world-wide license to publish or reproduce the published form of this manuscript, or allow others to do so, for the United States Government purposes. The Department of Energy will provide public access to these results of federally sponsored research in accordance with the DOE Public Access Plan.}

\begin{frontmatter}



\title{\textcolor{black}{Parallel} Quantum Computing Simulations via Quantum Accelerator Platform Virtualization}


\author[inst1]{Daniel Claudino}

\affiliation[inst1]{organization={Quantum Information Science Section, Oak Ridge National Laboratory},
            addressline={1 Bethel Valley Road}, 
            city={Oak Ridge},
            postcode={37831}, 
            state={TN},
            country={USA}}

\author[inst2]{Dmitry I. Lyakh}
\author[inst2]{Alexander J. McCaskey}

\affiliation[inst2]{organization={NVIDIA Corporation},
            city={Santa Clara},
            state={CA},
            country={USA}}

\begin{abstract}
Quantum circuit execution is the central task in quantum computation. Due to inherent quantum-mechanical constraints, quantum computing workflows often involve a considerable number of independent measurements over a large set of slightly different quantum circuits. Here we discuss a simple model for parallelizing simulation of such quantum circuit executions that is based on introducing a large array of virtual quantum processing units, mapped to classical HPC nodes, as a parallel quantum computing platform. Implemented within the XACC framework, the model can readily take advantage of its backend-agnostic features, enabling parallel quantum circuit execution over any target backend supported by XACC. We illustrate the performance of this approach by demonstrating strong scaling in two pertinent domain science problems, namely in computing the gradients for the multi-contracted variational quantum eigensolver and in data-driven quantum circuit learning, where we vary the number of qubits and the number of circuit layers. The latter (classical) simulation leverages the cuQuantum SDK library to run efficiently on GPU-accelerated HPC platforms.
\end{abstract}




\end{frontmatter}


\section{Introduction}

Quantum computing is largely an umbrella term that refers to several remarkably distinct architectural, logical, and computational approaches. However, this picture is dominated by the digital, gate-based model of quantum computation, and while there are theoretical guarantees that other competing modes are equally valid, such as adiabatic quantum computing,~\cite{Aharonov2008} those remain a much more niche technology. One trademark of the digital quantum computers is that, despite their striking difference from conventional computers, they retain the ability to enable some quantum operations to behave as analogs of conventional logic gates (for example, the classical \texttt{NOT} gate and the quantum \texttt{X} gate).

At the same time, quantum computers exhibit many unique characteristics that are inherently quantum, and an important example of which is quantum measurement. The state of a closed quantum system can be fully defined by its wave function, a mathematical object that has information about the behavior of the quantum system at every point in space. A critical complication arises from the fact that one is denied direct access to such an object, and is limited to drawing indirect conclusions based on quantum measurements. In turn, these measurements yield classical outcomes according to the corresponding observable operator through which the quantum state is measured, but which ends up affecting the original quantum state. This is a process known as the collapse of the wave function, which practically means that the quantum state, initially in a superposition of the eigenstates of the observable operator, upon measurement collapses into one of those eigenstates, revealing the probabilistic and destructive nature of the quantum measurement. Thus, in a general case where a quantum state is not prepared as an eigenstate of the quantum operator to be observed/measured, the measurement leads to a distribution of probabilities over its eigenstates. The prepared quantum state then needs to be resolved in such an eigenbasis, in which case many measurements need to take place in order to render a reliable statistics, each requiring a quantum state to be prepared ``from scratch''. \textcolor{black}{It is important to note the output information from this process is classical in nature. To illustrate this point, in the common case of digital quantum computers that operate on qubits which can store an arbitrary quantum superposition of two basis states, the outcome of measurements on each qubit is still classical, either 0 or 1. Thus, quantum algorithms ultimately still produce classical information.}

There are various ways the information in a quantum state can be encoded and manipulated, resulting in a multitude of simulation techniques whose inner workings follow those principles. The state can be naturally represented as a vector in an exponentially large Hilbert space, with the density matrix formalism requiring an even larger space, such that both can in principle be exact. Alternatively, there are much more computationally amenable representations of quantum states such as tensor network and stabilizer states which are applicable to certain classes of cases. Recognizing that each available mode of simulation has strengths and weaknesses means that one would ideally have a flexible framework that would easily grant access to this myriad of simulators while abstracting their intricacies. This is the foundational design argument in the XACC quantum-classical computing framework,~\cite{xacc1, xacc2} which puts forth a modular infrastructure for hardware-agnostic compilation and simulation of quantum circuits.

A common pitfall shared by all the different representations above is that general simulation of quantum states by classical means demands an exponential amount of resources, e.g., time and/or memory. Fortunately, this problem can be partially alleviated by turning to high-performance computing (HPC), which has become an invaluable tool in extending the scale and scope of how classical computers can aid in quantum computing research. An example of this is the validation of quantum protocols in which such simulations provide a setting where the deleterious effects of noise can be carefully modeled, or altogether removed from the computation. If, on the one hand, the nature of quantum measurements implies repeating seemingly the same tasks many times over, on the other hand, one can exploit the fact that all instances of these tasks are independent from one another, alluding to carrying out those measurements in an embarrassingly parallel fashion, which has become commonplace in HPC. This realization is one of the main motivations behind our proposed parallel quantum computing workflow.

In terms of programming and execution models, quantum computing has displayed a similar trajectory to that of accelerated computing based on the graphical processing units (GPU). That is, both approaches imply a heterogeneous computing system of which some parts serve as accelerators called to handle a limited subset of computational tasks that play to their strengths. While this paradigm has already proven extremely successful, heterogeneity increases complexity in various levels of the stack. This can be offset by adopting suitable abstractions that model the interaction between heterogeneous components, one example of which is compute resource virtualization \cite{SUNDERAM19991699}. Similar to the powerful quantum-classical programming framework \texttt{CUDA-Q} from NVIDIA \cite{cudaq}, in our approach each compute node (CPU or GPU) is {\color{black}modeled after} a \textit{virtual quantum processing unit} (QPU). {\color{black}In other words, QPUs are simulated as quantum virtual machines that are executed by classical compute resources, i.e., CPUs and GPUs.} In this picture, we leverage the standard message passing interface (MPI) orchestrate classical communication between virtual QPUs, thus being able to compose the entire quantum-classical parallel workflows.

{\color{black}One of the main technological limitations in quantum computing refers to the narrow time window during which assertions regarding the state prepared by a quantum computer can be made reliably, which is referred to as coherence time and which prevents quantum circuits from implementing a large number of gates. In various applications, this can be circumvented at the expense of an increasing number of circuit repetitions. This trade-off can be exploited in a variety of ways, and is at the heart of many quantum algorithms and subroutines, which is one of the strongest arguments in favor of a parallel or distributed quantum computing workflow. One of the most promising use cases of quantum computing, and which was the motivating application behind Feynman's argument,\cite{Feynman1982} is the simulation of physical systems, with the leading algorithm being the variational quantum eigensolver.\cite{vqe} This algorithm alleviates the circuit depth requirement of more precise alternatives (i.e. quantum phase estimation) by performing classical optimization of quantum circuit parameters, where the search for optimal parameters is informed by the classical outcomes of a much larger number of measurements. Similarly, circuit complexity can be lowered both in terms of the depth and width (number of qubits) by increasing the number of measurements.\cite{PhysRevX.6.021043, Steudtner2018} One prominent example is that of circuit cutting techniques,\cite{PhysRevLett.125.150504,Tuysuz2023,harrow2024optimal} which can lower the circuit depth requirements by cutting circuits into smaller, simpler ones, which incurs an exponential measurement overhead.}

{\color{black}Against this backdrop, the design rationale behind XACC not only embodies the versatility required to ensure performance across these different modalities, granted in part by its robust plugin infrastructure, but also attains a high degree of interoperability due to its layered compilation approach. Higher-level quantum programming languages, and even domain-specific languages, can be parsed into the XACC intermediate representation (IR). Developers then target backend-specific languages with compiler plugins that map the XACC IR onto the instruction set of the desired backend. Within this paradigm, users can ``mix and match'' languages and backends, while programming in their preferred languages and are freed from needing to possess explicit knowledge regarding compilation and corresponding code optimizations. Looking forward, this means that XACC is already equipped with the main features required to manage hybrid, highly interoperable workflows, where computations carried out by quantum accelerators are embedded in a largely classical computational pipeline.}

While the most {\color{black}immediate} application of the proposed execution model would be the {\color{black} parallel simulation of quantum circuits with classical numerical simulators}, as discussed above, {\color{black} over quantum processors would also be attractive and fully supported by our approach{\color{black}, akin to the ensemble quantum computing model.\cite{PhysRevA.69.052303}} With this in mind, our main contributions are:

\begin{itemize}
    \item We virtualize quantum compute resources, which we call \textit{virtual QPUs}, by abstracting away specifics of desired  numerical simulators and actual quantum hardware;
    \item Provide a flexible and robust communication framework based on the message passing interface (MPI) standard;
    \item Demonstrate our approach's scalability via important domain science problems deploying simulations using both CPUs and GPUs, in two different HPC systems, and with distinct numerical simulator backends.
\end{itemize}

The rest of the paper is organized as follows. We first review previous related work in Section \ref{sec:related}. In Section \ref{sec:design} we present the most relevant design elements of our virtualization scheme and some basic aspects of our implementation. In Section \ref{sec:performance} we discuss the performance and scalability of our approach in light of two relevant problems in the domain of quantum simulation and quantum machine learning. Finally, in Section \ref{sec:conclusion} we summarize our main findings and consider potential future ramifications.

{\color{black}
\section{Related Work}
\label{sec:related}

The execution of an exponential number of calculations in parallel is at the heart of the expected speedups in various instances of quantum computing. This was realized early in the development of quantum computing as natural execution model to be employed in many quantum algorithms, the quintessential example being the Shor algorithm for prime number factorization. Later on, parallel execution in quantum computing gained prominence primarily as a means to circumvent the exponential demand in memory/time for classical simulation of quantum systems. The latter modality is where our platform falls into and the one for which we will discuss similar instances in the literature. 

The first example of a parallel simulator in the literature comes from the work of Obenland and Despain, which was implemented in C and enabled parallelization via MPI. A distinct feature of this simulator is its implementation of quantum gates in terms of experimental gate parameters in order to mimic laser pulses in trapped ions.\cite{obenland1998parallel} Another example of a simulator implemented in C is found in Ref.\citenum{PhysRevA.66.062317}, however with parallelization based on a multi-threaded shared memory model. Additionally, the implementation of a massively parallel simulator, initially implemented in Fortran 90 and parallelized via MPI, has undergone several updates and offers both shared and distributed memory models.\cite{De_Raedt2007,De_Raedt2019}

With the emergence of graphical processing units, their intrinsic architectural paradigm can be naturally leveraged for parallel execution over their many threads, and here we emphasize the initial works of Gutierrez et al.\cite{Gutierrez2008} and Amariutei and Caraiman,\cite{6085728} demonstrating parallelization within a single GPU.

 Recent works present simulators designed to be deployed in the HPC environments. Two scalable, CPU-only simulators\cite{Li2020,Wang2021} have demonstrated performance for quantum chemistry applications~\cite{10.5555/3571885.3571903,morita2024simulator}. The NWQ suite of simulators provides CPU and GPU implementations of the state vector and density matrix representations of the quantum state.\cite{li2020density,li2021svsim,10.1145/3624062.3624221,suh2024simulating} It is also worth mentioning works in specialized simulators, with the ones based on tensor networks standing out, such as ExaTN,\cite{Lyakh2022} QTensor,\cite{Lykov_2021} and ITensor.\cite{itensor}

 Perhaps the closest examples to our virtualization platform is CUDA-Q\cite{cudaq} and QuEST,\cite{jones2019quest} later renamed to the Intel Quantum Simulator, and currently named the Intel Quantum SDK. They offer both shared (OpenMP) and distributed (MPI) memory parallelization and can target both CPU and GPU backends. However, the Intel Quantum SDK differs from XACC in that the set of backends made available by the Intel Quantum SDK is currently restricted to classical numerical simulators, while XACC and CUDA-Q encompass those as well as actual quantum hardware.


}
\section{Design and Implementation}
\label{sec:design}

\subsection{Design}
\label{ssec:design}

In XACC, quantum hardware is viewed as accelerators embedded in a classical computational workflow. This model draws inspiration from HPC nodes and how they leverage GPUs, which is considered a staple of heterogeneous computation. Within this picture, two important abstractions are in order, namely those facilitated by the \texttt{Accelerator} and \texttt{AcceleratorBuffer} interfaces. The former enables access to a target backend to carry out quantum circuit executions. The \texttt{AcceleratorBuffer} serves two main purposes: it models the underlying register of quantum bits (qubits) and provides data structures to persist information resulting from circuit execution, such as bit-string counts, and hardware specifics metadata like error rates. From a more practical standpoint, XACC defers circuit execution to concrete implementations of the \texttt{Accelerator} interface, which can be numerical simulators (IBM's Aer, Google's qsim\cite{quantum_ai_team_and_collaborators_2020_4023103}, Quantum\texttt{++}\cite{qpp}, TNQVM,\cite{tnqvm1, tnqvm2} etc.) or quantum computing vendor hardware (IBM, Rigetti, Quantinuum, etc.).

{\color{black}One of the most distinctive features of XACC is it intermediate representation (IR), which enables richly composable quantum workflows. A variety of programming languages can be parsed into the XACC IR as well as circuit construction via a builder class. At that point, the XACC puts forth a \texttt{Compiler} interface whose main task is to map the instructions in the XACC IR onto the native gate set of the desired backend. Once compiled into the set of instructions and the format expected by the backend of choice, the circuits are offloaded for execution. The primary result of circuit execution is the distribution of observed bitstrings over the possible $2^N$ outcomes, which may in turn be further processed in order to provide answers according to the computational task at hand. The main components of this workflow, which also serves as a basis for understanding the parallel pipeline, are illustrated in Figure \ref{fig:xacc_workflow}.}

\begin{figure}[ht!]
    \centering
    \includegraphics[width=\columnwidth]{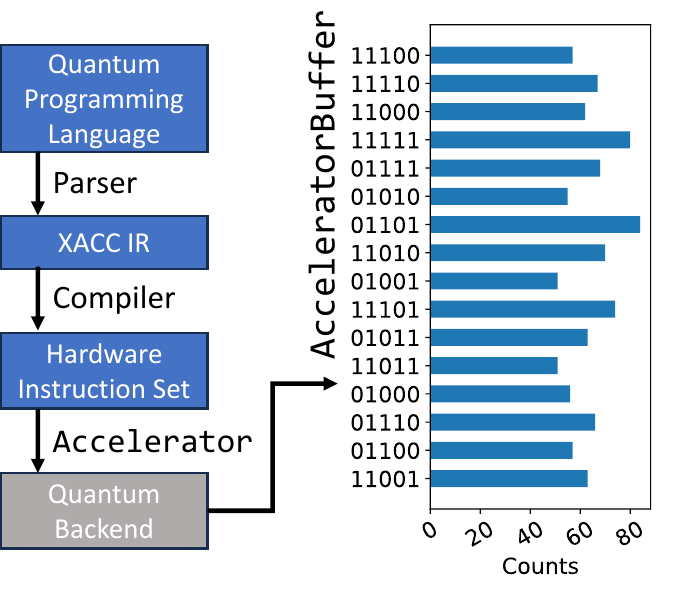}
    \caption{Summary of the XACC circuit execution workflow.}
    \label{fig:xacc_workflow}
\end{figure}

The salient features of the design behind our virtualization scheme are summarized in the diagram in Figure \ref{fig:uml}, which follow the usual decorator pattern. While the \texttt{Accelerator} interface prescribes other methods, we suppress them here because they are not relevant in the current context. Here we highlight that{\color{black}, for a collection of virtual QPUs over which we intend to distribute/parallelize the computational workload,} each virtual QPU {\color{black}is assigned ownership of and has access} to an independent instance of the chosen \texttt{Accelerator}, with the two concrete implementations used in this work being illustrated in Figure \ref{fig:uml}, namely \texttt{qsim} and Aer. Also of importance is that all the necessary communication is facilitated by \texttt{qpuComm} attribute in each virtual QPU, which will be discussed in more detail later on.

\begin{figure*}[htbp!]
  \includegraphics[width=\textwidth]{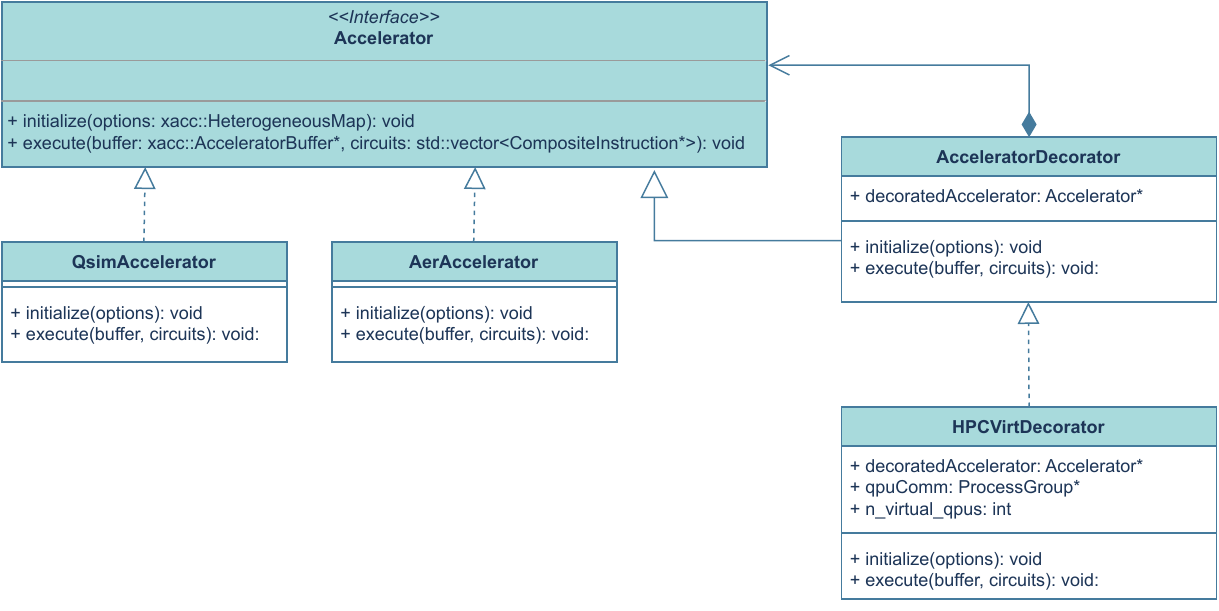}
  \caption{\label{fig:uml}The class diagram representing the design of the \texttt{HPCVirtDecorator}.}
\end{figure*}

Before we delve into the details of our implementation, we consider appropriate to discuss some of the relevant features of MPI in light of the programming and execution model we propose here. In this context, we judge essential to avail ourselves of this standard to enable communication and synchronization among virtual QPUs, which has a proven track-record of maturity and robustness in HPC environments. One aspect of MPI that is crucial to our model is the concept of communicators, which represent groups of processes for MPI provide extensive capabilities that enables efficient and streamlined communication. {\color{black}The communication functionality relevant to the current work is encapsulated by \texttt{ProcessGroup} (Figure \ref{fig:uml}), which is a wrapper to native MPI calls (in \texttt{C}) with the intention of simplifying message passing and communication management.} In our framework, each virtual QPU is modeled after an MPI communicator, allowing processes associated with the same virtual QPU to exchange data and coordinate execution. Another important concept is MPI collective communication operations, of which we highlight \texttt{MPI\_Bcast()} and \texttt{MPI\_Allgatherv()}, which are used for broadcasting data from one process to all others or gathering data from all processes, respectively. These operations are employed to consolidate execution results obtained from individual virtual QPUs that become accessible to all processes. Because several MPI routines are \textit{non-blocking} in the most general sense of the term, synchronization is ensured by \texttt{MPI\_Barrier()}. By leveraging MPI's communication primitives, our framework achieves efficient and scalable parallel execution of quantum circuits across heterogeneous computing resources.

In order to facilitate the discussion, we establish that data / operations done at the virtual QPU level are referred to as \textit{local}, while those that take place across/are shared by all virtual QPUs are \textit{global}. Once {\color{black} the \texttt{Accelerator} offloads the quantum circuits to the quantum backend of choice (Figure \ref{fig:xacc_workflow})}, barring trivial cases where the number of virtual QPUs is either set to 1 or is less than the number of available MPI processes, all available MPI processes are assigned to a certain virtual QPUs and identified by a \textit{color}. 
{\color{black}Furthermore, each process associated with a given color has access to the corresponding \texttt{qpuComm} object (Figure \ref{fig:uml}). We assigned one process under each \texttt{qpuComm}, conventionally the one with rank 0, to be responsible for managing all the necessary communication. In a similar way as the MPI processes are grouped according to colors, each virtual QPU is assigned a subset of the entire collection of quantum circuits to be executed. A local \texttt{AcceleratorBuffer} is instantiated to collect and persist the outcomes of execution called by the corresponding virtual QPU and offloaded to the quantum backend of choice. At this point, each virtual QPU is equipped with all the required ingredients to carry out the required local circuit executions, whose outcomes are generally probability distributions, but can also be expectation values (of type \texttt{double}), in the case of numerical simulators. These results are broadcast in order to be consolidated under an \texttt{AcceleratorBuffer} that is shared among all virtual QPUs, thus enabling all MPI processes to have access to all the execution data. This is summarized in Figure \ref{fig:nodes}.}

\begin{figure}[htb!]
    \centering
    \includegraphics[width=\columnwidth]{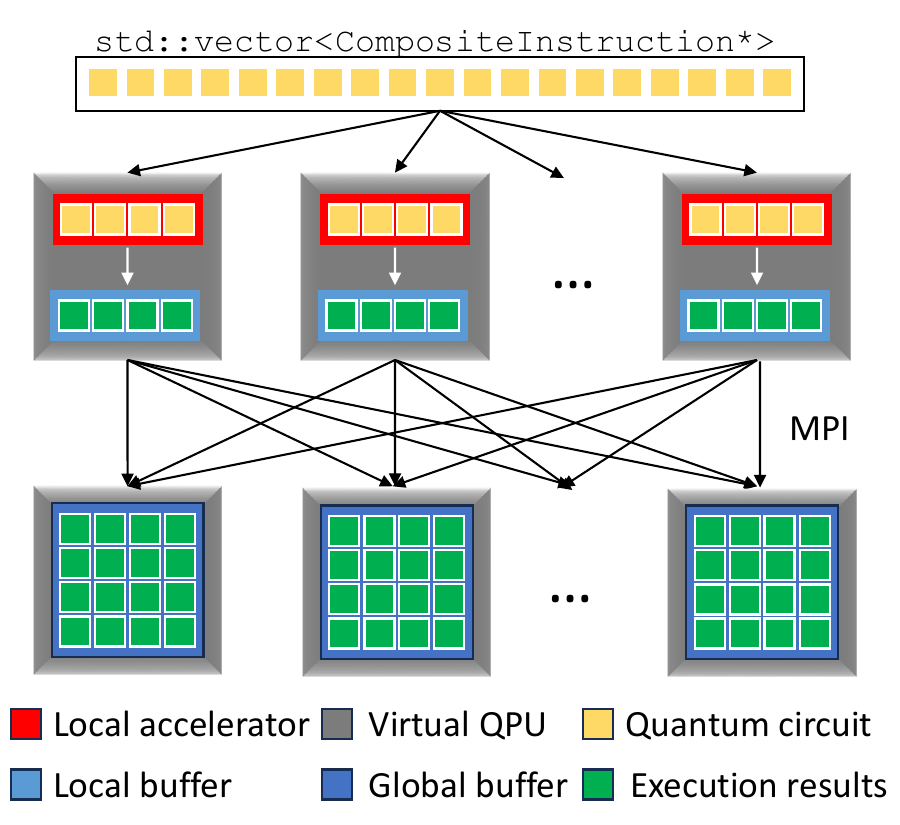}
    \caption{Main data structures and communication pattern across virtual QPUs. The white arrows represent a call to {\color{black}the \texttt{Accelerator} instance in each virtual QPU to offload quantum circuits to the backend,} and the two rows of virtual QPUs represent different states of the same virtual QPUs before (top) and after (bottom) communication.}
    \label{fig:nodes}
\end{figure}

\subsection{Implementation in the XACC framework}


{\color{black}In this section we discuss some pertinent details of the virtualization platform in the XACC framework, which is facilitated by turning to the example in Figure  \ref{fig:execute} that illustrates its main features. We bring attention to the fact that XACC is entirely developed in modern C\texttt{++}.}

\begin{figure}[htbp]
\begin{lstlisting}[numbers=left]
using namespace xacc;
// get reference to Accelerator
auto accelerator = getAccelerator("qsim");
// allocate a register of nBits qubits
buffer = xacc::qalloc(nBits);
// execute std::vector<CompositeInstruction*>
// in serial manner
accelerator->execute(buffer, circuits);
// now turn to parallel execution
// options is a HeterogeneousMap data structure 
// that maps n-virtual-qpus to # of virtual QPUs
auto options = {{"n-virtual-qpus", 256}};
// decorate Accelerator
accelerator = getAcceleratorDecorator(
      "hpc-virtualization", 
      accelerator, options);
// decorated accelerator executes circuits
accelerator->execute(buffer, circuits);
\end{lstlisting}
\caption{\label{fig:execute}Code snippet showcasing how to execute a collection of circuits in XACC using the \texttt{qsim}\cite{quantum_ai_team_and_collaborators_2020_4023103} simulator and how to enable parallel execution via the \texttt{HPCVirtDecorator}.}
\end{figure}

In order to tie back to the previous section, the \texttt{Accelerator} interface enables access to a target backend via the \texttt{execute()} method, which takes as arguments a quantum circuit (or an \texttt{std::vector} thereof) and a \texttt{AcceleratorBuffer$^*$}, whose function is examined in Section \ref{ssec:design}. In XACC, a quantum circuit is embodied by an instance of \texttt{CompositeInstruction}, which provides the abstraction for a cohesive set \texttt{Instruction}s (e.g. quantum gates). A collection of \texttt{CompositeInstruction}s can be executed via \texttt{Accelerator::execute()}, as shown in Figure \ref{fig:execute}.


Normally, the call to \texttt{Accelerator::execute()} would serialize the execution by iterating over the contents of the \texttt{\\std::vector<CompositeInstruction$^*$>} (line 8 in Figure \ref{fig:execute}). In order to enable parallel quantum circuit execution, the original \texttt{Accelerator::execute()} functionality is augmented based on the decorator pattern (\texttt{HPCVirtDecorator}). We note that XACC makes extensive use of decorators as a very convenient means to abstract complex pre- and post-processing protocols of quantum circuit execution data. The usage of the proposed \texttt{HPCVirtDecorator} is illustrated in Figure \ref{fig:execute}, lines 12-18. The \texttt{options} object (line 12) is a $\texttt{HeterogenousMap}$, which is a flexible data structure that can map \texttt{string}s to any value type (via \texttt{std::any<T>}), and through which one passes various parameters to customize execution. Among these parameters, which the implementation elects to be required or optional, we illustrate a map with just \texttt{n-virtual-qpus} by which we pass to \texttt{HPCVirtDecorator} the number of virtual QPUs (256 in our example in Figure \ref{fig:execute}, line 12). This parameter's default value is 1, in which case the trivial, serial circuit execution ensues.

In passing, we point that the \texttt{getAcceleratorDecorator} wrapper takes the name of the decorator, and the \texttt{Accelerator} instance to be decorated, and a \texttt{HeterogeneousMap} with parameters, e.g., number of virtual QPUs. Parallel circuit execution follows by retrieving \texttt{HPCVirtDecorator} from the XACC service registry via the string identifier \texttt{hpc-virtualization} and calling its \texttt{execute()} method. Moreover, the \texttt{xacc} namespace provides the \texttt{getAccelerator} and the \texttt{\\getAcceleratorDecorator} wrappers (lines 8 and 14, respectively) that obviates the use of templated functions explicitly.

As mentioned above, all the communication across virtual QPUs is handled by the message passing interface (MPI) standard. With the XACC framework being entirely developed in C\texttt{++}, we are faced with choice of either linking the\\\texttt{HPCVirtDecorator} against the Boost implementation of MPI or to carry out all the MPI calls in C. We chose the latter, as MPICH or OpenMPI are both adherent to the MPI specification and updates tend to be more timely. That being said, that modeling of each virtual QPU after an MPI communicator is facilitated by a set of helper functions that wrap the relevant C calls under \texttt{ProcessGroup} and thus enable utilization of C\texttt{++} smart pointers {\color{black}(\texttt{qpuComm} in Figure \ref{fig:uml})}.


At this point, the results from execution in each virtual QPU need to consolidated, which in principle could be achieved via broadcast ($\texttt{MPI\_Bcast()}$). However, the results of execution of quantum circuits on digital quantum computers are akin to a hash table of bit-strings and the number of times they occur (counts), which in C\texttt{++} can be simply embodied by the \texttt{std::map} data structure, with no analog in C. This is a drawback of opting to C MPI calls instead of relying on the Boost implementation. Each local buffer (\texttt{AcceleratorBuffer}), when handed over to \texttt{Accelerator::execute()} stores execution information of each unique circuit in a \textit{child} buffer. For our purposes, these children buffer persists three basic pieces of execution information that we are concerned with: 1) a unique \texttt{std::string} identifier, whose name defaults to the memory address the child buffer points to and is often overridden by the string representation of the observable being measured; 2) a hash table of bit-strings to counts, that is, \texttt{std::map<std::string, int>}; 3) (Optional) the expectation value associated with the measurement, which is relevant in certain applications and is of type \texttt{double}.

We chose to circumvent the limitations {\color{black}arising from turning to a C implementation of MPI} by expressing all the information in the children buffers in terms of MPI native data types. In other words, this means using \texttt{MPI\_INT}, \texttt{MPI\_DOUBLE}, and \texttt{MPI\_CHAR} in lieu of \texttt{int}, \texttt{double}, and \texttt{std::string}, respectively. Apart from the communication involved in sending and receiving execution data, this also entails explicit knowledge of the amount of data being communicated. Once the necessary communication is carried out across the virtual QPUs, each MPI communicator has access to all execution data and then are in position to rebuild the global buffer (\texttt{buffer} in line 18) as though the quantum circuits were executed serially. This is illustrated by the arrows and the second row in Figure \ref{fig:nodes}. 


Finally, we comment on an implementation detail that may be relevant depending on the intended application. By the end of \texttt{HPCVirtDecorator::execute()} call, we impose synchronization across all virtual QPUs and the global buffer is now accessible by all processes. In this case, the global buffer differs from that of trivial execution by storing the corresponding MPI rank of the process. This is important because in several contexts the post-processing that follows from circuit execution can itself be data and compute intensive, thus by enforcing post-processing to take place only from the data associated with a certain rank, e.g., 0, we can avoid redundant computations and data transfer. 


\section{Performance evaluation}
\label{sec:performance}

\subsection{Multi-contracted variational quantum eigensolver}
\label{ssec:mcvqe}

Many applications of quantum computing revolve around estimating the expectation value of some quantum mechanical observable. The prime example of this is the estimation of the energy eigenspectrum of a physical system, which revolves around measurements of the quantum circuit via the Hamiltonian operator $\hat{H}$, with emphasis in the ground state energy $E_0$. In the context of quantum computing, the most natural representation of such operators is in terms of Pauli spin operators. For complex quantum systems, the Hamiltonian can be comprised of a large number of tensor products of such Pauli spin operators, e.g., $\mathcal{O}(N^4)$ for the Hamiltonian of a collection of interacting electrons in a molecule. Each of these terms in $\hat{H}$ requires a quantum circuit to be measured, an  undertaking that can greatly benefit from a parallel execution model.  

A proposed means to circumvent current quantum hardware limitations (e.g., circuit depth) in ground state energy estimation is the ubiquitous variational quantum eigensolver (VQE) algorithm.\cite{vqe} Briefly, one choose a circuit implemented by the unitary $\hat{U}(\vec{\theta})$ where $\vec{\theta}$ is a series of parameters, generally single-qubit rotation gate angles, that are varied until a cost function, which here is the expectation value of $\hat{H}$, is minimized. This is summarized in Equation \ref{eq:vqe} 
\begin{equation}
    \label{eq:vqe}
    E_0 \leq \langle \hat{H} \rangle (\vec{\theta}) = \min_{\vec{\theta}} \langle 0| \hat{U}^\dagger(\vec{\theta}) \hat{H} \hat{U}(\vec{\theta}) |0 \rangle.
\end{equation}

The multi-contracted variant of the VQE algorithm (MC-VQE)\cite{mcvqe} specializes in the absorption spectrum of a collection of weakly-interacting ``monomers'', mapping their ground and first excited states to the $|0\rangle$ and $|1\rangle$ states of a qubit. The salient features of this model can be successfully captured by the Hamiltonian $\hat{H}$ in Equation \ref{eq:aiem}

\begin{align}
    \hat{H} &= \mathcal{E} + \sum_\text{A}\mathcal{X}_\text{A}\hat{X}_{\text{A}}+\mathcal{Z}_\text{A}\hat{Z}_{\text{A}}  \nonumber \\ 
    &+ \sum_\text{A, B}\mathcal{XX}_\text{AB}\hat{X}_{\text{A}}\hat{X}_{\text{B}} + \mathcal{XZ}_\text{AB}\hat{X}_{\text{A}}\hat{Z}_{\text{B}} \nonumber \\
    &+ \mathcal{ZX}_\text{AB}\hat{Z}_{\text{A}}\hat{X}_{\text{B}} + \mathcal{ZZ}_\text{AB}\hat{Z}_{\text{A}}\hat{Z}_{\text{B}}, \label{eq:aiem}
\end{align}
where $\hat{X}$ and $\hat{Z}$ are the Pauli X and Z spin operators, and the coefficients, e.g., $\mathcal{X}_\text{A}$, are real numbers.

The target state builds upon a classical configuration interaction singles approximation ($\psi_\text{CIS}$).\cite{Foresman1992} At this point, the qubit register holds a W state generalized to the desired number of qubits.\cite{PhysRevA.62.062314} Entanglement necessary to go beyond this approximation is introduced via entanglers $\hat{U}_\text{AB}(\vec{\theta})$ that are comprised of 6 single-qubit $R_y$ rotation gates and act between qubits associated with adjacent monomers via two CNOT gates. The circuit and entangler gate layouts are shown in Figure \ref{fig:entangler}.

\begin{figure}[htbp!]
    \centerline{\includegraphics[width=.75\columnwidth]{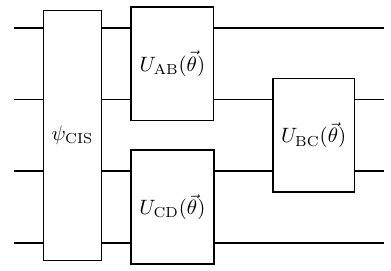}}
    \centerline{\includegraphics[width=\columnwidth]{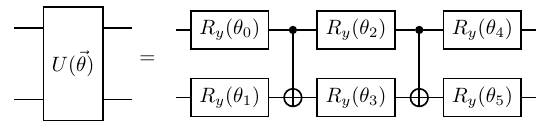}}
    \caption{Layouts of the circuit and chosen entangler used in the MC-VQE algorithm.}
    \label{fig:entangler}
\end{figure}

Each monomer has its own $\psi_\text{CIS}$ state preparation step, but they all share the same entanglers $\hat{U}_\text{AB}(\vec{\theta})$. The central task in the MC-VQE algorithm is to minimize the energy average over all monomers, by minimizing the cost function in Equation \ref{eq:vqe} with the Hamiltonian in \ref{eq:aiem}. Because $\psi_\text{CIS}$ is often a good first approximation and the final energy is often downhill from the energy of that initial state, it is desirable to leverage gradient-based optimization. This can readily benefit from application of the parameter-shift rule\cite{parameter_shift}, which in this case involves $\mathcal{O}(2N_HN_\theta)$ circuit executions per parameter update, where $N_H$ is the number of terms in the Hamiltonian (Equation \ref{eq:aiem}), and $N_\theta$ is the size of $\vec{\theta}$. Important metrics related to the parameter update in this algorithm for a range of qubits are displayed in Table \ref{tab:mcvqe_simulations}. It is important to note that here we take advantage of the fact that $R_y(\theta)R_y(\phi) = R_y(\theta+\phi)$ to merge adjacent $R_y$ gates, thus reducing the number of variational parameters to be optimized.

\begin{table}[htbp!]
\caption{Number of monomers/qubits ($N_q$), number of terms in the Hamiltonian $N_H$, number of variational parameters $N_\theta$, and number of circuits in a single parameter update.}
\begin{center}
\begin{tabular}{|c|c|c|c|}
\hline
$N_q$ & $N_H$ & $N_\theta$ & Number of circuit executions \\
\hline
16 & 92 & 76 & 13984 \\
\hline
18 & 104 & 86 & 17888 \\
\hline
20 & 116 & 96 & 22272 \\
\hline
22 & 128 & 106 & 27136\\
\hline
\end{tabular}
\label{tab:mcvqe_simulations}
\end{center}
\end{table}

Our scheme is implemented in C\texttt{++} as is the XACC framework. Python users can also take advantage of this functionality, along with much of what XACC has to offer, via bindings facilitated by Pybind11\cite{pybind11}. We investigate the strong scaling related to the computation of MC-VQE gradients using the parameter-shift rule where \texttt{qsim}\cite{quantum_ai_team_and_collaborators_2020_4023103} serves as the XACC virtual backend (numerical, noiseless statevector simulator). In this study, each virtual QPU comprises of a compute node in the Oak Ridge Computing Facility (OLCF) Andes system, each with 2 AMD EPYC 7302 16 Core Processor 3.0 GHz, totalling 32 cores per node. Scaling plots based on the parameters in Table \ref{tab:mcvqe_simulations} are presented in Figure \ref{fig:plots} with runtimes from both native C\texttt{++} executables and also relying on the corresponding Python bindings.

\begin{figure}[ht!]
    \centerline{\includegraphics[height=0.9\textheight]{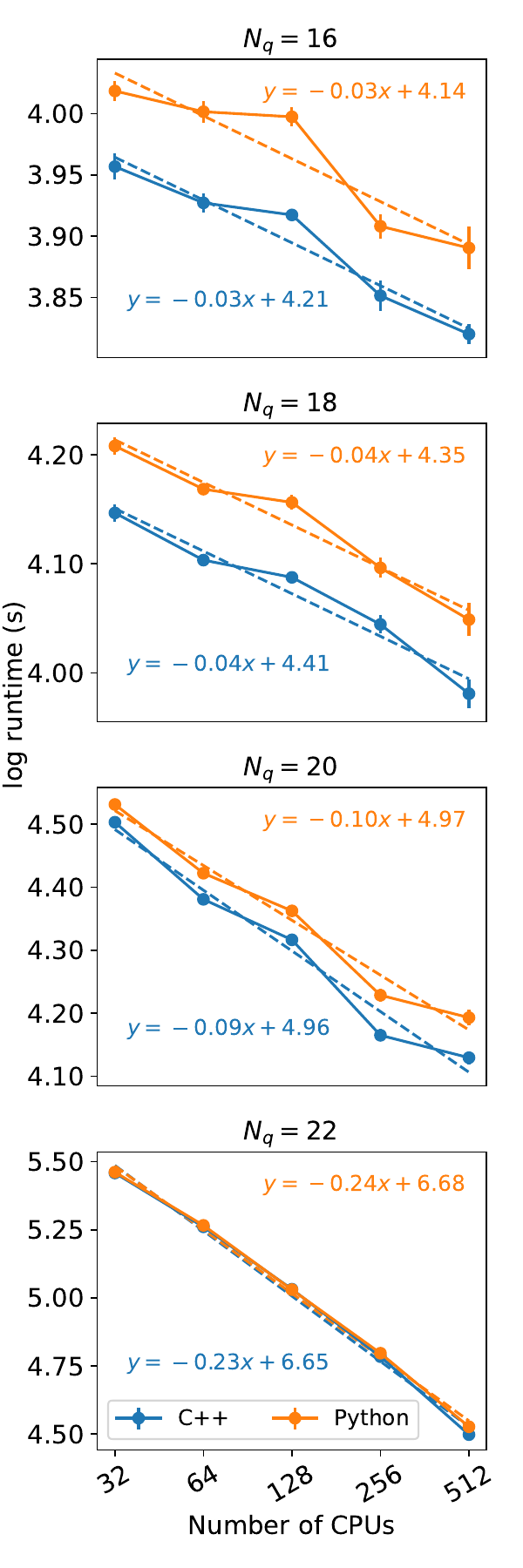}}
    \caption{Strong scaling plots of the MC-VQE gradients for 16-22 qubits for a single parameter update.}
    \label{fig:plots}
\end{figure}

As the number of qubits increase, together with number of circuits, we observed a trend that consistently better approximates the expected linear behaviour. There is a small but noticeable disadvantage when relying on the Python interface in comparison with exclusive C\texttt{++}, even though the overall qualitative pattern emerges in either case, pointing to the introduction of some overhead in the Python case. This overhead is becomes negligible as the number of qubits increase, pointing to the fact the runtime in this case is largely dominated by the execution of more demanding circuits.

{\color{black}In light of the discussion in Section \ref{sec:design}, the \texttt{vector} containing the circuits in Table \ref{tab:mcvqe_simulations} (third column) is split into as many segments as the desired number of virtual QPUs, here determined by the $x$-axis in Figure \ref{fig:plots} and which are in turn offloaded to its instance of the \texttt{qsim} simulator. For a given choice of the circuit parameters $\bar{\theta}$ in Equation \ref{eq:vqe} and Figure \ref{fig:entangler}, and casting Equation \ref{eq:aiem} as a sum of generic terms $\hat{P}_i$ weighed by scalars $c_i$, the final outcome of quantum circuit execution in this case is to estimate the expectation values $\langle 0| \hat{U}^\dagger(\bar{\theta}) \hat{P}_i \hat{U}(\bar{\theta}) |0 \rangle$, which are scalars. While in reality this requires each expectation value to be resolved in the eigenspectrum of the corresponding $\hat{P}_i$, a numerical simulator can directly evaluate such a scalar, which we do here. Thus, each local buffer stores a collection of children buffers, each with the string representation of the $\hat{P}_i$ it measured and the resulting expectation value.

The local information now needs to be consolidated. We assigned the rank-0 MPI process in each qpuComm communicator to handle the necessary communication. The local information, i.e., expectation values and buffer names, are gathered (via \texttt{MPI\_Allgather()} and \texttt{MPI\_Allgatherv()}) into \texttt{vector}s of the respective types, which in turn each \texttt{qpuComm} broadcasts to its subordinate process (via \texttt{MPI\_Bcast()}). At this point all MPI processes have access to all the execution data, which are used to populate the parent buffer that was originally passed to the \texttt{execute()}. The copy of the parent buffer managed by each process also stores its MPI rank (\texttt{int}) as metadata, and synchronization across all communicators is enforced prior to the end of the \texttt{execute()} call.
}

\subsection{Data-driven circuit learning}


The use case presented in Subsection \ref{ssec:mcvqe} is primarily concerned with the computation of expectation values, which in the case of execution on a quantum circuit simulator can be evaluated directly. More generally, the output of execution of quantum circuits on digital quantum hardware is the probability distribution over all possible $2^N$ bit-strings, which can give insight into the prepared quantum state and can, in turn, be used to estimate expectation values.

Various applications of quantum algorithms revolve around how well one can match a target probability distribution. One of such approaches is the data-driven circuit learning algorithm (DDCL).\cite{Benedetti2019} The goal is to learn a set of rotation angles $\vec{\theta}$ that parameterize a quantum circuit such that the output distribution $P_\theta$ minimizes the Jensen–Shannon (JS) divergence\cite{https://doi.org/10.48550/arxiv.1901.08047, PhysRevD.106.096006} for a given target distribution $P$ as shown in Equation \ref{eq:js}

\begin{equation}
    \label{eq:js}
    \text{JS}(P|P_\theta) = \frac{1}{2}\sum_b\left( P(b)\log \frac{P(b)}{M} + P_\theta(b)\log \frac{P_\theta(b)}{M}\right),
\end{equation}
with $M=(P+P_\theta)/2$ and $P(b)$ being the probability ascribed to the bit-string $b$ in the distribution $P$.

Since $P_\theta$ can be estimated by measuring all qubits in the $Z$-basis, it involves the measurement of a single Pauli string, and the total cost per gradient estimate scales as $\mathcal{O}(2N_\theta)$. The circuit starts with each pair of neighboring qubits being entangled as a Bell pair. This is followed by layers of gates where the angles $\vec{\theta}$ appear in the layout portrayed in Figure \ref{fig:qml_entangler}, partially inspired by Ref. \citenum{PhysRevD.106.096006}.

\begin{figure}[htbp!]
    \centerline{\includegraphics[width=.75\columnwidth]{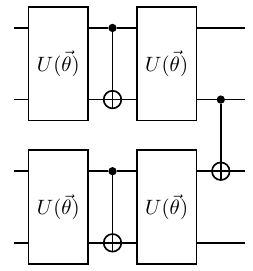}}
    \centerline{\includegraphics[width=\columnwidth]{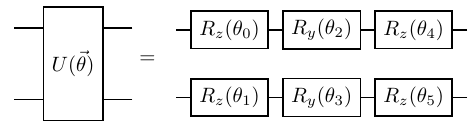}}
    \caption{Layout of the layer used in the DDCL algorithm.}
    \label{fig:qml_entangler}
\end{figure}

\begin{table}[htbp]
\caption{Number of qubits ($N_q$) and number of variational parameters $N_\theta$. The number of circuits implementations in a single parameter update is $2N_\theta$.}
\begin{center}
\begin{tabular}{|c|c|c|}
\hline
$N_q$ & $N_\theta$ & Number of circuit executions \\
\hline
20 & 1200 & 2400 \\
\hline
22 & 1320 & 2640\\
\hline
24 & 1440 & 2880  \\
\hline
26 & 1560 & 3120 \\
\hline
\end{tabular}
\label{tab:qml_simulations}
\end{center}
\end{table}

We again monitor the time associated with a single parameter update, whose runtime is largely dominated by the execution of the $2N_\theta$ quantum circuits required by the parameter-shift rule. In this case, our virtual QPUs are powered by a locally modified version of the IBM Aer simulator which leverages the efficient \texttt{cuStateVec} library available in the \texttt{cuQuantum SDK} from NVIDIA \cite{cuQuantumSDK}. Simulations were performed on the OLCF Summit HPC system, where each virtual QPU maps onto a single NVIDIA Tesla V100 GPU out of each compute node, with the results reported in Figure \ref{fig:qml_plot}.

\begin{figure}
    \centering
    \includegraphics[width=\columnwidth]{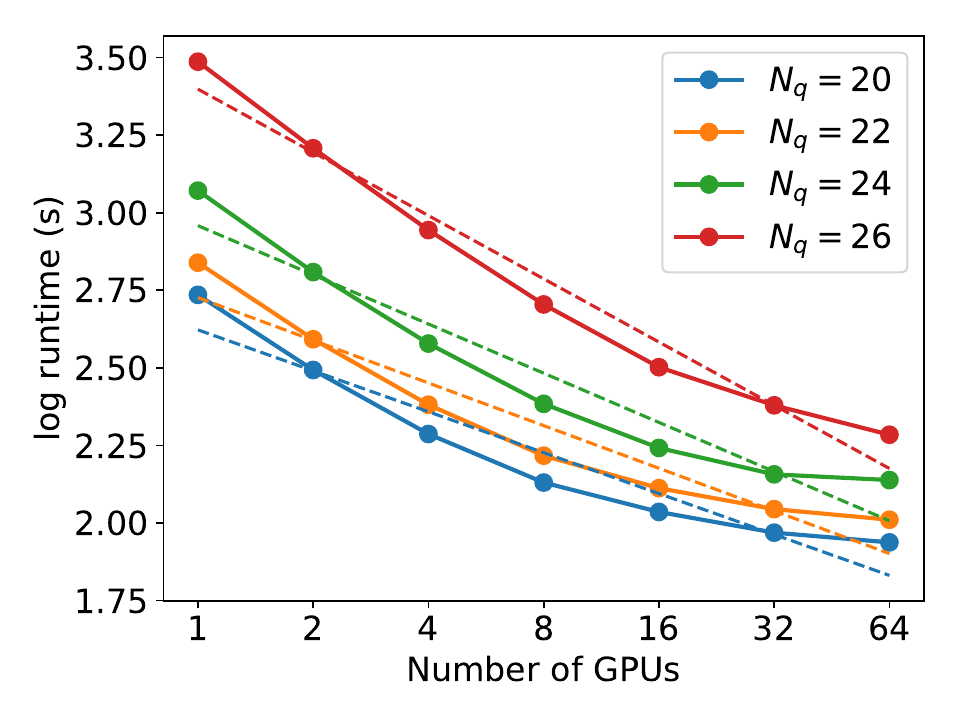}
    \caption{Strong scaling plots of the DDCL gradients for a single parameter update with 20-26 qubits for the circuit in Figure \ref{fig:entangler} with 10 layers.}
    \label{fig:qml_plot}
\end{figure}

{\color{black}
In Figure \ref{fig:qml_plot} we plot the average runtimes over 10 independent gradient evaluations, along with error bars given by the standard deviation ($\sigma$). It so happens that the standard deviation is negligible in the scale of the plot, and not visible in Figure \ref{fig:qml_plot}. In Table \ref{tab:qml_stat} we report those values, along with the equation of the line from the linear fits from the corresponding data points.
}
\begin{table}[htbp]
\caption{\color{black}Number of qubits ($N_q$), maximum standard deviation (max $\sigma$) and the equation corresponding to the linear fits of log$_{10}$(runtime) vs. log$_2$ (number of GPUs) shown in Figure \ref{fig:qml_plot}.}
\begin{center}
{\color{black}
\begin{tabular}{|c|c|c|}
\hline
$N_q$ & max $\sigma$ (s) & Linear fit \\
\hline
20 & 1.49 & $y = -0.132x + 2.622$ \\
\hline
22 & 0.97 & $y = -0.137x + 2.726$ \\
\hline
24 & 0.77 & $y = -0.159x + 2.958$ \\
\hline
26 & 2.96 & $y = -0.204x + 3.398$ \\
\hline
\end{tabular}
}
\label{tab:qml_stat}
\end{center}
\end{table}

{\color{black}In the case of GPU execution of the DDCL algorithm, whose performance is illustrated by Figure \ref{fig:qml_plot}, we notice that the strong scaling deviates from the linear trend as the number of GPUs increase. This can be attributed to the fact that with a larger number of GPUs, each GPU execute fewer circuits, and the time overhead associated with data transfer between CPU and GPU eventually becomes a considerable part of the overall runtime.}

{\color{black}The discussion on the messages exchanged in the previous example (end of Section \ref{ssec:mcvqe}) largely holds in the case of the present use case, with one crucial difference. Because the DDCL algorithm is concerned with matching a target distribution, the relevant outcome of circuit execution in this case are the bitstring counts, not expectation values. Since XACC stores those as a hash table (\texttt{std::map<std::string, int>}), and such a data structure finds analog in C, we start by storing the observed bitstrings and the corresponding counts in appropriate local \texttt{vector}s. The communication pattern involving those follow closely that regarding the other \texttt{vector}s considered in Section \ref{ssec:mcvqe}.}

\section{Conclusions}
\label{sec:conclusion}

We put forward a parallel quantum computing software tool that is based on virtualization of compute resources by abstracting details of the execution and backend away from the end user. It takes advantage of the foundational architectural feature that grants XACC a high degree of backend agnosticism to be well positioned to stay relevant in the rapidly evolving field of quantum computing. The independent nature of quantum circuit executions can be explored by a parallel workflow, which we accomplish by extending the capabilities found in the \texttt{Accelerator} interface following a decorator design pattern and availing ourselves of the acclaimed MPI standard. We highlight the performance of our scheme in the OLCF Andes and Summit leadership computers, with both CPU-only and GPU-only runs, in the context the resource-intensive gradient evaluation tasks in some representative quantum algorithms. The strong scalings drawn from these simulations provide evidence that our proposal constitute a sound and promising strategy in the pursue of significant speedups in quantum computing workflows that revolved around execution of a large number of circuits.

One of the main reasons our approach stands out is due to the modularity afforded by XACC. An example of this is that by virtualizing compute resources, the approach presented here is not constrained to classical simulation of quantum circuits, but is equally applicable to workflows where the accelerators are actual quantum hardware, and in principle even a collection of radically different hardware technologies. This is line with ensemble quantum computing,\cite{PhysRevA.69.052303} with a previous demonstration presented in Ref. \citenum{eqc}.

{\color{black}Finally, in the present work we only consider a parallel workflow of execution of quantum programs, where each virtual QPU independently executes an entire program, which corresponds to a quantum circuit. A major development in scalability would be achieved by an analog of MPI that would enable quantum communication, making possible large scale quantum computation using an array of several smaller quantum computers~\cite{qmpi}. This scheme would allow for distributed quantum computing by having this collection of QPUs execute a single program, in which each individual QPU is assigned a subset of the instructions, e.g. gates, that compose the quantum circuit in question.\cite{qdc, qpram} There is currently a specification for such a communication protocol, but no available implementation, and interested readers are referred to extensive reviews on the topic.\cite{caleffi2022distributed,barral2024review} As we anticipate this new paradigm in light of what XACC is capable of, minor changes may required in order to accommodate the communication primitives proposed by QMPI, such as ensuring that all operations are reversible, which assumes uncomputation, as well as modifications and/or additions in terms of abstraction layers that will enable the virtualization platform to carry out distributed quantum computing. An immediate example is how to determine which QPU implements a specific set of gates. We are confident that XACC's flexible design will carry over and be crucial in moving toward this mode of quantum computing and is well positioned to embrace those advancements.}

\section{Acknowledgments}

DC and DL acknowledge funding by the US Department of Energy award ERKCG13 provided by the Office of Basic Energy Sciences. AJM acknowledges funding by the US Department of Energy Office of Science Advanced Scientific Computing Research (ASCR), Accelerated Research in Quantum Computing (ARQC). This research used resources of the Oak Ridge Leadership Computing Facility, which is a DOE Office of Science User Facility supported under Contract DE-AC05-00OR22725.
\\~\\
\textbf{Declaration of generative AI and AI-assisted technologies in the writing process}

During the preparation of this work the author(s) used ChatGPT in order to improve readability. After using this tool/service, the author(s) reviewed and edited the content as needed and take(s) full responsibility for the content of the publication.



 \bibliographystyle{elsarticle-num} 
 \bibliography{cas-refs}





\end{document}